\documentclass[aps, prb, preprint, superscriptaddress, twocolumn, 10pt]{revtex4-1}

\usepackage{amsmath}
\usepackage{amsfonts}
\usepackage{amssymb}
\usepackage{graphicx}
\usepackage{multirow}

\begin{document}

\title{Dynamic pathway of the photoinduced phase transition of TbMnO$_3$}
\date{\today}

\author{Elisabeth Bothschafter}
\email{elisabeth.bothschafter@outlook.com}
\affiliation{Swiss Light Source, Paul Scherrer Institut, 5232 Villigen, Switzerland}

\author{Elsa Abreu}
\email{elsabreu@phys.ethz.ch}
\affiliation{Institute for Quantum Electronics, Eidgen\"ossische Technische Hochschule (ETH) Z\"urich, 8093 Z\"urich, Switzerland}

\author{Laurenz Rettig}
\affiliation{Swiss Light Source, Paul Scherrer Institut, 5232 Villigen, Switzerland}
\affiliation{Current address: Department of Physical Chemistry, Fritz-Haber-Institut of the Max Planck Society, Berlin 14915, Germany}

\author{Teresa Kubacka}
\affiliation{Institute for Quantum Electronics, Eidgen\"ossische Technische Hochschule (ETH) Z\"urich, 8093 Z\"urich, Switzerland}

\author{Sergii Parchenko}
\affiliation{Swiss Light Source, Paul Scherrer Institut, 5232 Villigen, Switzerland}

\author{Michael Porer}
\affiliation{Swiss Light Source, Paul Scherrer Institut, 5232 Villigen, Switzerland}

\author{Christian Dornes}
\affiliation{Institute for Quantum Electronics, Eidgen\"ossische Technische Hochschule (ETH) Z\"urich, 8093 Z\"urich, Switzerland}

\author{Yoav William  Windsor}
\affiliation{Swiss Light Source, Paul Scherrer Institut, 5232 Villigen, Switzerland}
\affiliation{Current address: Department of Physical Chemistry, Fritz-Haber-Institut of the Max Planck Society, Berlin 14915, Germany}

\author{Mahesh Ramakrishnan}
\affiliation{Swiss Light Source, Paul Scherrer Institut, 5232 Villigen, Switzerland}

\author{Aurora Alberca}
\affiliation{Swiss Light Source, Paul Scherrer Institut, 5232 Villigen, Switzerland}

\author{Sebastian Manz}
\affiliation{Department of Materials, Eidgen\"ossische Technische Hochschule (ETH) Z\"urich, 8093 Z\"urich, Switzerland}

\author{Jonathan Saari}
\affiliation{Institute for Quantum Electronics, Eidgen\"ossische Technische Hochschule (ETH) Z\"urich, 8093 Z\"urich, Switzerland}

\author{Seyed M. Koohpayeh}
\affiliation{Institute for Quantum Matter (IQM), Department of Physics and Astronomy, Johns Hopkins University, Baltimore, MD 21218, USA}

\author{Manfred Fiebig}
\affiliation{Department of Materials, Eidgen\"ossische Technische Hochschule (ETH) Z\"urich, 8093 Z\"urich, Switzerland}

\author{Thomas Forrest}
\affiliation{Diamond Light Source, Harwell Science and Innovation Campus, Didcot, Oxfordshire OX11 0DE, United Kingdom}

\author{Philipp Werner}
\affiliation{Department of Physics, University of Fribourg, 1700 Fribourg, Switzerland}

\author{Sarnjeet S. Dhesi}
\affiliation{Diamond Light Source, Harwell Science and Innovation Campus, Didcot, Oxfordshire OX11 0DE, United Kingdom}

\author{Steven L. Johnson}
\email{johnson@phys.ethz.ch}
\affiliation{Institute for Quantum Electronics, Eidgen\"ossische Technische Hochschule (ETH) Z\"urich, 8093 Z\"urich, Switzerland}

\author{Urs Staub}
\email{urs.staub@psi.ch}
\affiliation{Swiss Light Source, Paul Scherrer Institut, 5232 Villigen, Switzerland}

\begin{abstract}
We investigate the demagnetization dynamics of the cycloidal and sinusoidal phases of multiferroic TbMnO$_3$ by means of time-resolved resonant soft x-ray diffraction following excitation by an optical pump.
Using orthogonal linear x-ray polarizations, we suceeded in disentangling the response of the multiferroic cycloidal spin order from the sinusoidal antiferromagnetic order in the time domain.
This enables us to identify the transient magnetic phase created by intense photoexcitation of the electrons and subsequent heating of the spin system on a picosecond timescale.
The transient phase is shown to be a spin density wave, as in the adiabatic case, which nevertheless retains the wave vector of the cycloidal long range order.
Two different pump photon energies, 1.55 eV and 3.1 eV, lead to population of the conduction band predominantly via intersite $d$-$d$ transitions or intrasite $p$-$d$ transitions, respectively.
We find that the nature of the optical excitation does not play an important role in determining the dynamics of magnetic order melting.
Further, we observe that the orbital reconstruction, which is induced by the spin ordering, disappears on a timescale comparable to that of the cycloidal order, attesting to a direct coupling between magnetic and orbital orders.
Our observations are discussed in the context of recent theoretical models of demagnetization dynamics in strongly correlated systems, revealing the potential of this type of measurement as a benchmark for such complex theoretical studies.

DOI:
\end{abstract}

\maketitle

\newpage

\section{Introduction}

Multiferroics of type-II exhibit magnetic order that induces spontaneous electric polarization with a strong magnetoelectric coupling \cite{Fiebig2016}.
This coupling holds potential for novel electronic and magnetic applications such as magnetic storage that could be manipulated by the electric field of strongly focused light pulses.
Depending on the response time to an optical excitation, it may be possible to control the orientation of magnetoelectric domains with laser pulses on picosecond timescales.
Such an ultrafast polarization switching has been predicted in single-phase multiferroic TbMnO$_3$ when applying very strong THz frequency electromagnetic field pulses. \cite{Mochizuki2010}

TbMnO$_3$ crystalizes in a perovskite structure with \emph{Pbnm} symmetry.
Below T$_{N_1} \simeq 42$ K it exhibits an antiferromagnetic sinusoidal spin density wave order.
Upon further cooling, these collinearly oriented spins acquire an additional magnetic component.
Cycloidal order arises within the bc-plane, propagating along the b-axis and slightly elliptically deformed.
This cycloid is incommensurate with the lattice and gives rise to a spontaneous ferroelectric polarization along the c-axis below T$_{N_2} \simeq 27$ K, the multiferroic phase transition temperature \cite{Kimura2003, Matsubara2015}.
The polarization direction is defined by the rotational sense of the spin cycloid \cite{Yamasaki2007}.
This long-range spin order gives rise to distinct magnetic Bragg reflections, with wave vectors (0 $q$ 0), (0 $q$ 1) and (0 1-$q$ 1) \cite{Kenzelmann2005}, and to higher order quadrupole-like reflections due to orbital reconstruction, with wave vector (0 2$q$ 0) \cite{Lovesey2013}.
These have been investigated in detail by neutron scattering \cite{Kenzelmann2005, Senff2007, Kajimoto2004}, and by non resonant \cite{Fabrizi2009, Walker2011, Walker2013} and resonant x-ray Bragg diffraction \cite{Lovesey2013, Walker2009, Wilkins2009, Forrest2008, Strempfer2008, Mannix2007}.

Understanding the dynamics of the magnetic order is highly desirable, particularly in the context of ultrafast switching.
Ultrafast pump-probe resonant x-ray diffraction using optical or THz pump pulses and soft x-ray probe pulses can provide insight into the magnetic states of strongly correlated systems by accessing the complex dynamics that follow the interaction of intense light pulses with these magnetic states \cite{Kubacka2014, Johnson2015, Forst2011, DeJong2013, Caviglia2013, Forst2015, Boeglin2010}.
In the case of TbMnO$_3$, it has recently been shown that it is possible to drive coherent dynamics of the magnetic structure via resonant THz excitation of an electromagnon. \cite{Kubacka2014}
Actual switching of the ferroelectric polarization by such a resonant excitation would require much higher field strengths, but could be possible in the near future.
The effect of femtosecond near-infrared excitation ($h\nu_1$ = 1.55 eV) on the magnetic order of TbMnO$_3$ has also been investigated using resonant soft x-ray diffraction as a probe. \cite{Johnson2015}
That study reports the magnetic order to be suppressed on a timescale of several tens of picoseconds, which is very slow compared to the sub-picosecond demagnetization of other antiferromagnetically ordered systems such as CuO \cite{Johnson2012}.
The reason for the stability of the antiferromagnetic state in TbMnO$_3$ has not yet been established.
Optical control over the antiferromagnetic order via the secondary ferroelectric polarization has
been demonstrated in a recent study by Manz et al. \cite{Manz2016}
In that work, reversible switching of the antiferromagnetic domain orientation is achieved in a two-color laser experiment.
Laser beams heat the sample across the multiferroic ordering temperature T$_{N_2}$, after which the depolarization field of the re-cooling sample reverses the electric polarization and, with it, the primary antiferromagnetic order.
Bidirectional switching is achieved by taking advantage of the difference in optical and thermal penetration depths for the two laser colors.

In light of these recent time-resolved and static results, it is important to investigate further the dynamics of the demagnetization process in TbMnO$_3$.
Understanding the nature of the transient states and of the hierarchy of interactions that define them, and understanding how the transient states compare to the equilibrium phases of the system and how they are affected by the photoexcitation pathway is essential in order to achieve a precise control of the magnetic order in this prototypical multiferroic material.
In turn, control over the processes and timescales involved in magnetic order dynamics are at the core of any ultrafast switching application.
One important question for TbMnO$_3$ lies in the origin of the slow demagnetization process, and how this bottleneck relates to the properties of the initial and transient states, and to the excitation pathway that connects them.

In the present study, we use time- and polarization-dependent resonant soft x-ray diffraction at the Mn $L_2$ edge to characterize the transient magnetic state that develops after optical excitation of the multiferroic and sinusoidal phases with picosecond-scale resolution.
We compare the effect of different excitation pathways out of the cycloidal phase, and analyze the coupling between spin and orbital magnetic order.
First, we investigate whether femtosecond photoinduced heating of the multiferroic phase leads directly to the demagnetized state or whether the transition occurs via a transient spin-density wave state, as in the adiabatic thermal transition.
Second, we compare two pump energies ($h\nu_1$ = 1.55 eV and $h\nu_2$ = 3.1 eV) to address the question of whether the nature of the photoexcitation pathway plays a role in the dynamics of the demagnetization process.
We consider either intersite charge-transfer Mn $3d$ - Mn $3d$ transitions at the lower edge of the conduction band or higher lying intrasite O $2p$ - Mn $3d$ transitions \cite{Moskvin2010, Kim2006}.
Third, since the spin order is accompanied by orbital reconstruction as evidenced by the second order reflection (0 2$q$ 0), we investigate the dynamic coupling in magnetic and orbital order signals by comparing their decay times upon ultrafast laser excitation.
Finally, we discuss the results of recent calculations which predict slow magnetic dynamics due to long doublon-hole recombination times.
A model that can be quantitatively compared to TbMnO$_3$ will, however, have to include multi-orbital effects and electron-phonon coupling.
Our data will serve as a good benchmark for such a model.

\section{Static resonant soft x-ray diffraction}

We use a bulk single crystal sample of TbMnO$_3$, cut such that its surface is perpendicular to the b-axis.
Growth and characterization details are described elsewhere \cite{Lovesey2013}.
The sample was cut and polished down to a thickness of ~200 micrometers so as to improve the cooling of the surface probed by the x-rays.
After polishing, the sample was annealed at 650 degrees Celsius for $\sim$120 hours in air at atmospheric pressure.

Resonant soft x-ray diffraction at the Mn $L_2$ absorption edge ($\sim$653 eV) on b-cut TbMnO$_3$ crystals enables the observation of the first- and second-order reflections of the magnetic sinusoidal and cycloidal order, (0 $q$ 0) and (0 2$q$ 0).
The azimuthal orientation of the sample is chosen such that the c-axis is perpendicular and the a-axis is parallel to the scattering plane.
The linear polarization of the incoming light is set either parallel ($\pi$-polarization) or perpendicular ($\sigma$-polarization) to the scattering plane.
The polarization-dependent scattered signal intensity provides a clear distinction between sinusoidal and cycloidal magnetic phases, as only the latter shows a signal with $\sigma$-polarization at the (0 $q$ 0) reflection \cite{Lovesey2013}. 

Static experiments were performed using the soft x-ray resonant scattering endstation RESOXS \cite{Staub2008} at the Surfaces/Interfaces Microscopy beamline X11MA \cite{Flechsig2010} at the Swiss Light Source.

\begin{figure} [htb!]
\begin{center}
\includegraphics[width=0.4\textwidth,keepaspectratio=true]{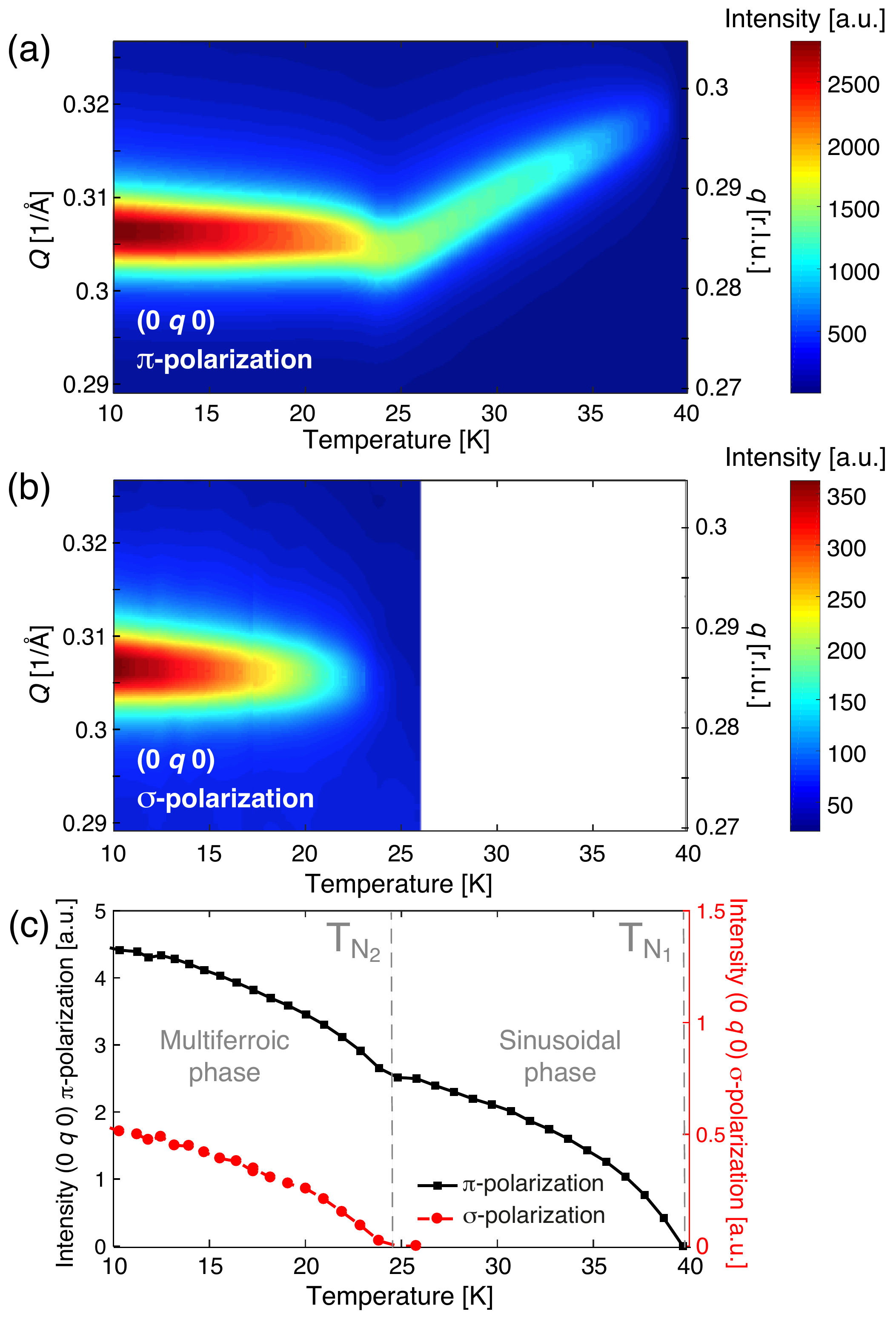}
\caption{
Intensity maps of the (0 $q$ 0) reflection recorded at 653 eV with incoming (a) $\pi$- and (b) $\sigma$-polarization as functions of temperature and momentum transfer (in 1/{\AA} on the left and in relative lattice units on the right, using the lattice constants from Blasco \emph{et al.} \cite{Blasco2000}).
Scattering occurs in the ab-plane for this azimuth of the sample.
(c) (0 $q$ 0) integrated intensity as a function of temperature for $\pi$- (black squares) and $\sigma$- (red circles) polarization.
Magnetic transition temperatures are shown as vertical dashed lines.
}
\label{static_q}
\end{center}
\end{figure}

Figure \ref{static_q} shows the momentum transfer and intensity of the (0 $q$ 0) magnetic reflection as functions of temperature and incident polarization.
Figures \ref{static_q}a and \ref{static_q}b are intensity maps vs. temperature and momentum transfer for $\pi$ and $\sigma$-polarized light, respectively.
Figure \ref{static_q}c plots the integrated intensity of the (0 $q$ 0) reflection vs. temperature for both x-ray polarizations.
The momentum transfer, which is related to the period of the spin-density wave modulation, is nearly constant up to $\sim$24 K, i.e. in the cycloidal phase.
Upon further warming, the momentum transfer of the $\pi$-polarized signal decreases slightly and then increases with temperature (Figs. \ref{static_q}a and \ref{static_q}c).
This behavior is associated with the sinusoidal phase, between about 24 K and 40 K \cite{Kenzelmann2005}.
The $\sigma$-polarized signal, sensitive only to the projection of spin correlations along the c-axis, is absent above 24 K (Figs. \ref{static_q}b and \ref{static_q}c).
Indeed, this projection is non-zero only for the non-collinear magnetic order that arises below T$_{N_2}$, such that the (0 $q$ 0)$_\sigma$ diffraction intensity serves as a signature of the multiferroic phase.

\begin{figure} [htb!]
\begin{center}
\includegraphics[width=0.4\textwidth,keepaspectratio=true]{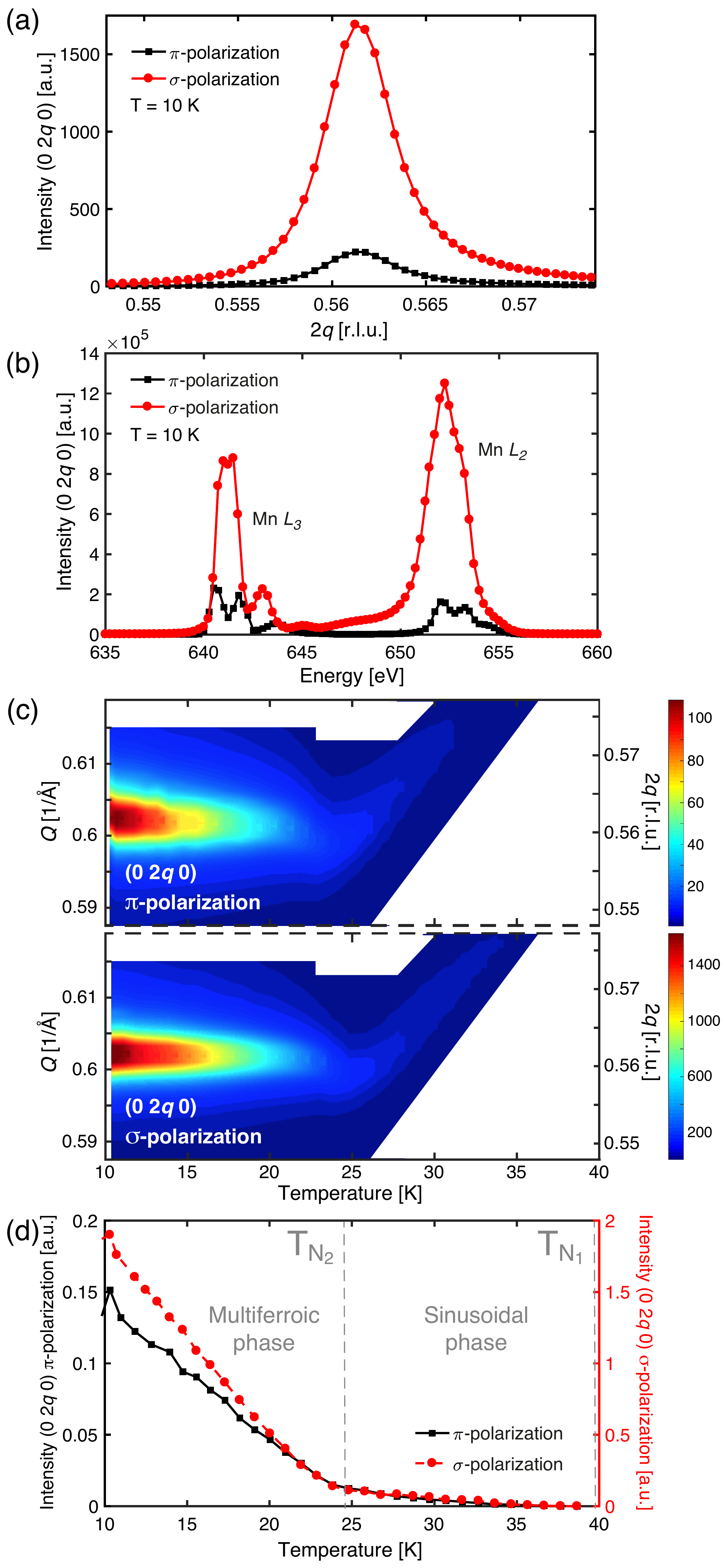}
\caption{
(0 2$q$ 0) intensity for $\pi$- (black squares) and $\sigma$- (red circles) polarization, as a function of (a) momentum transfer, at 10 K and 653 eV; (b) energy, at 10 K and at the peak position shown in (a); (d) temperature, at 653 eV (integrated intensity).
Magnetic transition temperatures are shown as vertical dashed lines in (d).
(c) Intensity maps of the (0 2$q$ 0) reflection recorded at 653 eV with incoming (top) $\pi$- and (bottom) $\sigma$-polarization as functions of temperature and momentum transfer (in 1/{\AA} on the left and in relative lattice units on the right, using the lattice constants from Blasco \emph{et al.} \cite{Blasco2000}).
Scattering occurs in the ab-plane for this azimuth of the sample. 
}
\label{static_2q}
\end{center}
\end{figure}

The second-order reflection, (0 2$q$ 0), can also be observed with resonant soft x-ray diffraction.
This reflection arises at resonance from the orbital reconstruction (a charge quadrupole) caused by magnetic order.
The additional charge-lattice coupling is expected to displace atoms which makes the reflection weakly observable off resonance \cite{Mannix2007}.
The momentum transfer and energy dependence of this reflection are shown in Figs. \ref{static_2q}a and \ref{static_2q}b, respectively, for both $\pi$- and $\sigma$-polarizations.
The large difference in spectral shape for $\pi$- and $\sigma$-polarizations supports the view that this reflection is of quadrupole electron density origin.
As evidenced by the temperature dependence shown in Figs. \ref{static_2q}c and \ref{static_2q}d, a clear increase in the (0 2$q$ 0) intensity for both $\pi$- and $\sigma$-polarization occurs with the onset of cycloidal order, upon cooling below 24K.
The momentum transfer of the (0 2$q$ 0) reflection for both polarizations is nearly constant in the cycloidal phase but increases slightly with temperature in the sinusoidal phase.
Similar to the (0 q 0) magnetic order signal (Fig. \ref{static_q}c), the (0 2q 0) orbital order signal also increases below $\sim$24 K, indicating an increase in the orbital reconstruction due to the cycloidal arrangement of spins.

\section{Photoinduced demagnetization dynamics -- Time-resolved resonant soft x-ray diffraction}

Time-resolved experiments were carried out at the beamline I06 of the Diamond Light Source, in low-alpha operation mode \cite{Martin2011}, using soft x-ray probe pulses with 9.4 ps or 15.5 ps duration (full width at half maximum, FWHM) and an energy tuned to the Mn $L_2$ absorption edge (Fig. \ref{static_2q}b).
Femtosecond optical pump pulses had $\sim$100 fs pulse duration (FWHM), 5 kHz repetition rate and variable fluence, and a photon energy of either $h\nu_1$ = 1.55 eV or $h\nu_2$ = 3.1 eV.

The scattered x-ray intensity was recorded with a multichannel plate detector that was time-gated to record in a 60 nanosecond time window around the temporal overlap with the pump laser, selecting a single pulse of the MHz low-alpha x-ray pulse train.
The acquired data was normalized by the reading from a picoammeter x-ray signal monitor, to correct for slow drifts in the x-ray pulse intensity.

In order to fulfill the diffraction condition, the incidence angle of both optical and x-ray pulses relative to the sample surface was set to 27$^\circ$ and 65$^\circ$ for the (0 $q$ 0) and (0 2$q$ 0) reflections, respectively.
The nearly collinear x-ray probe and laser pump pulses where focused to spot sizes of ~100 $\mu$m and ~150 $\mu$m diameter (FWHM), respectively.
The choice of laser spot size was a compromise between the need to minimize average heat load and the desire to pump the region probed with x-rays as homogeneously as possible.
The penetration depth of the 1.55 eV and 3 eV pumps are about 200 nm and 110 nm, respectively, with little dependence on the incidence angle.
The x-ray penetration depth is $\sim 35$ nm at 27$^\circ$ incidence and $\sim 70$ nm at 65$^\circ$ incidence \cite{Windsor2016}.
The probe penetration depth is therefore always smaller than the pump depth in the configurations used in this work.

All time-resolved data presented in the following are peak intensity values measured at $q \sim 0.286$, for (0 $q$ 0) (Fig. \ref{static_q}), and at $2q \sim 0.561$, for (0 $2q$ 0) (Fig. \ref{static_2q}).
Previous experiments have reported that the peak width does not vary upon photoexcitation \cite{Johnson2015}, and our static characterization shows it to remain mostly unchanged as temperature is varied below T$_{N_1}$ (not shown).
Our time-resolved data can therefore be compared to the static integrated intensities shown in Figs. \ref{static_q}c and \ref{static_2q}d.
The reported uncertainties in the data (error bars) correspond to the standard error of the mean.

We evaluated best fits to a function corresponding to a single exponential convolved with the x-ray pulse duration and a nonlinear least-squares fitting algorithm to extract the decay time constant of the photoinduced demagnetization.
Uncertainties on the fitted values of the decay time correspond to 95\% confidence intervals.
The fitting range is as shown in Figs. \ref{q_pisig}-\ref{2q}.
Temporal overlap between the laser pump pulse and the x-ray probe pulse was determined experimentally from data obtained at higher fluence (not shown), which show a strong reduction of the diffraction signal.
This experimentally determined value was adjusted during the fitting procedure, as a global fitting parameter for all datasets.

\subsection{Polarization-dependent measurement of the transient magnetic phase}

\begin{figure} [htb!]
\begin{center}
\includegraphics[width=0.4\textwidth,keepaspectratio=true]{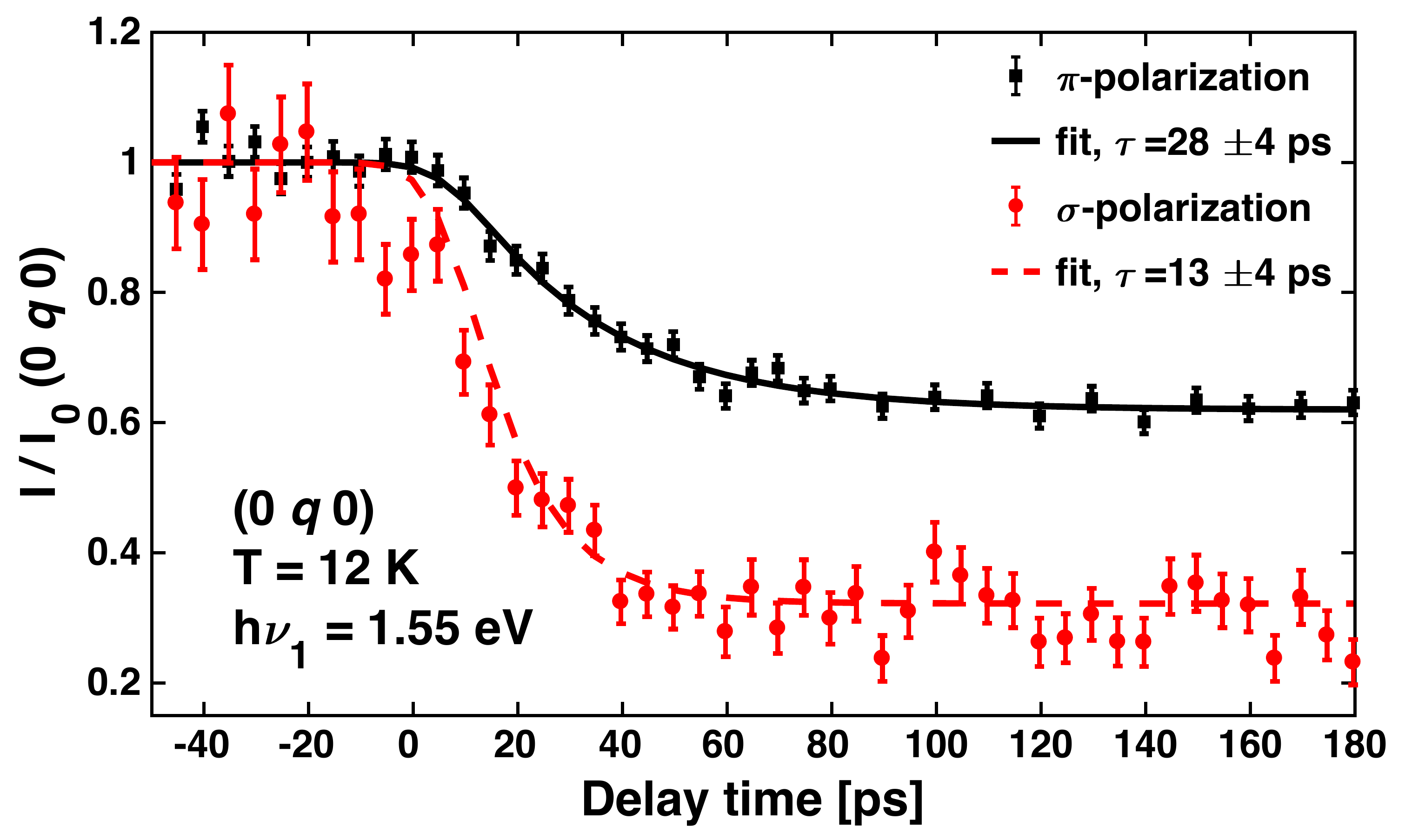}
\caption{
Time-resolved change in diffracted intensity of the (0 $q$ 0) reflection recorded at 12 K at the Mn $L_2$ edge for different x-ray probe polarizations, following photoexcitation at 1.55 eV with 1.3 mJ/cm$^2$ absorbed fluence.
$\pi$-polarization data and fit are shown as black squares and a full black line.
$\sigma$-polarization data and fit are shown as red circles and a dashed red line.
}
\label{q_pisig}
\end{center}
\end{figure}

Fig. \ref{q_pisig} presents the time-dependent change in diffraction intensity of the magnetic (0 $q$ 0) reflection upon excitation with a femtosecond laser pulse at 1.55 eV, for both $\sigma$- (red circles) and $\pi$- (black squares) x-ray polarization and for a 15.5 ps x-ray pulse duration (FWHM).
The absorbed fluence, accounting for reflectivity losses using optical data from Bastjan \emph{et al.} \cite{Bastjan2008}, was 1.3 mJ/cm$^2$ and the measured sample temperature was 12 K, well within the cycloidal phase of the material.
A single exponential fit to the data yields a decay time of 13$\pm$4 ps for $\sigma$- and 28$\pm$4 ps for $\pi$-polarization.
The timescale for melting of the long range order we observe here with $\pi$-polarization is slightly longer than the 22.3$\pm$1.1 ps decay time found by Johnson \emph{et al.} \cite{Johnson2015}.
Several reasons could contribute to this small difference.
First, the effective initial temperature of the probed region might be slightly higher than the measured sample temperature due to a possible larger average heating by the laser pulses impinging at a 5 kHz repetition rate compared to the 60 Hz repetition rate of the experiment of Johnson \emph{et al.} \cite{Johnson2015}.
Second, the length of the delay time trace extends here to 180 ps after photoexcitation compared to only 60 ps in the previous experiment \cite{Johnson2015}.
Restricting the current fitting range to 60 ps yields a 26$\pm$11 ps decay time, closer to the 22.3$\pm$1.1 ps value determined by Johnson \emph{et al.} \cite{Johnson2015} but with larger uncertainties. 
An influence of the fitting time window might be an indication of multiple decay constants, as has been observed e.g. for antiferromagnetic Ho \cite{Rettig2016}.
Third, another source of uncertainty in the evaluation of the decay time comes from the time-resolution of the experiment.
The latter is determined by the x-ray pulse shape and duration, which in the experiment presented here was 15.5 ps (FWHM), derived from the low-alpha x-ray bunch length and close to the relevant timescales of the experiment, vs. 100 fs (FWHM) for the setup used by Johnson \emph{et al.} \cite{Johnson2015}.
Finally, any inhomogeneity in the photoexcitation of the probed region is likely different between both experiments, leading to yet another source of discrepancy.
Given these experimental distinctions, the agreement between the timescale we observe (28$\pm$4 ps) and the one previously reported (22.3$\pm$1.1 ps \cite{Johnson2015}) is sufficient for a comparison of the conclusions from both experiments.

\begin{figure} [htb!]
\begin{center}
\includegraphics[width=0.4\textwidth,keepaspectratio=true]{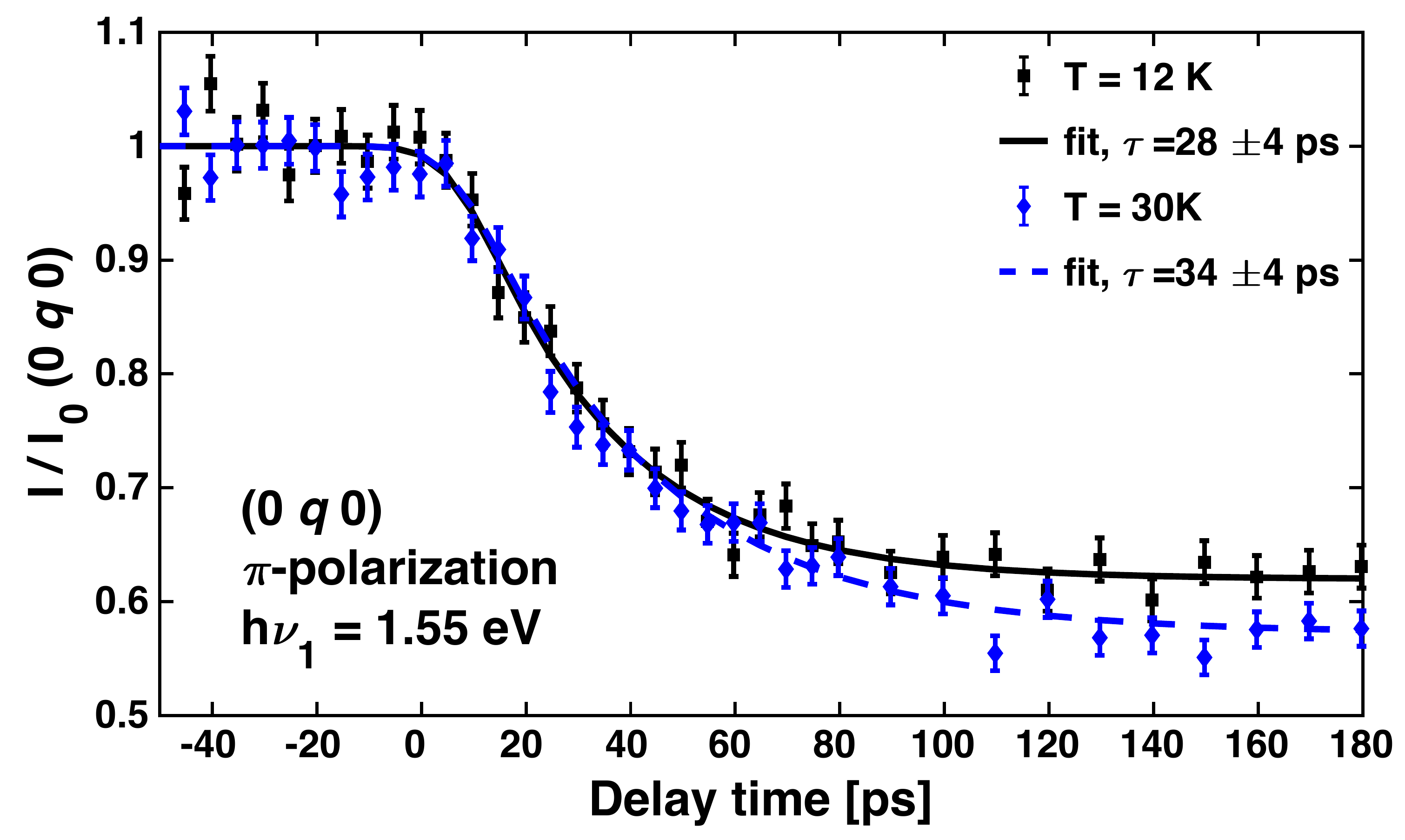}
\caption{
Time-resolved change in diffracted intensity of the (0 $q$ 0) reflection recorded at the Mn $L_2$ edge for $\pi$-polarization of the x-ray probe, following photoexcitation at 1.55 eV.
12 K data and fit are shown as black squares and a full black line.
30 K data and fit are shown as blue diamonds and a dashed blue line.
The absorbed fluence was 1.3 mJ/cm$^2$ at 12 K and 0.85 mJ/cm$^2$ at 30 K, chosen to achieve a similar maximum decrease in intensity.
}
\label{q_12K30K}
\end{center}
\end{figure}

Our main observation here is that the diffraction intensity of the (0 $q$ 0)$_\sigma$ channel, which is sensitive to the projection of the spin order along the c-direction, decreases significantly more relative to its equilibrium position than that of the (0 $q$ 0)$_\pi$ channel following photoexcitation of the multiferroic phase.
As seen in Fig. \ref{q_pisig}, the (0 $q$ 0)$_\sigma$ intensity drops by 70\% whereas the (0 $q$ 0)$_\pi$ intensity drops by only 40\%, i.e. 1.8 times less.
Decay times obtained from the fits to the $\sigma$- and $\pi$-polarized signals are 13$\pm$4 ps and 28$\pm$4 ps, respectively.

In order to compare these results to the demagnetization dynamics in the sinusoidal phase, we repeat the measurement at a sample temperature of 30K, which is above T$_{N_2}$ but below T$_{N_1}$.
As discussed above, only the $\pi$-polarization signal remains non-zero in this phase.
The pump fluence was reduced from 1.3 mJ/cm$^2$ to 0.85 mJ/cm$^2$ to match approximately the maximum transient decrease in $\pi$-polarized scattered intensity at 30 K to the one at 12 K (40\%).
Indeed, we know that there is a roughly linear relationship between the equilibrium scattered intensity and temperature (cf. Fig. \ref{static_q}c and Ref. \cite{Johnson2015}), and that the demagnetization dynamics are independent of the excitation fluence for transient changes below ~80\% \cite{Johnson2015}.
In order to compare the timescales at different temperatures we chose to match the total decrease in scattered intensity instead of matching the excitation density.
Dynamics of the (0 $q$ 0)$_\pi$ channel at 30 K are shown in Fig. \ref{q_12K30K} (blue diamonds), along with those at 12 K (black squares), for comparison.
The intensity drop is comparable for both temperatures, around 40\%, and fitting yields a decay time of 34$\pm$4 ps at 30 K.

\subsection{Influence of the pump photon energy on the demagnetization dynamics}

In our experimental configuration we can directly compare the effect of 1.55 eV vs. 3.1 eV pump photon energy on the timescale and magnitude of the transient demagnetization signal.
A 9.4 ps x-ray pulse duration (FWHM) was used for the $h\nu_2$ = 3.1 eV measurements.
In order to achieve a similar excitation density for photon energies of 3.1 eV and 1.55 eV, the decrease in penetration depth with increasing energy must be taken into account, in addition to reflectivity losses \cite{Bastjan2008}.
A reduced incident power at 3.1 eV, corresponding to a fluence of 0.8 mJ/cm$^2$, matches the absorbed energy density obtained with 1.3 mJ/cm$^2$ at 1.55 eV. 

\begin{figure} [htb!]
\begin{center}
\includegraphics[width=0.4\textwidth,keepaspectratio=true]{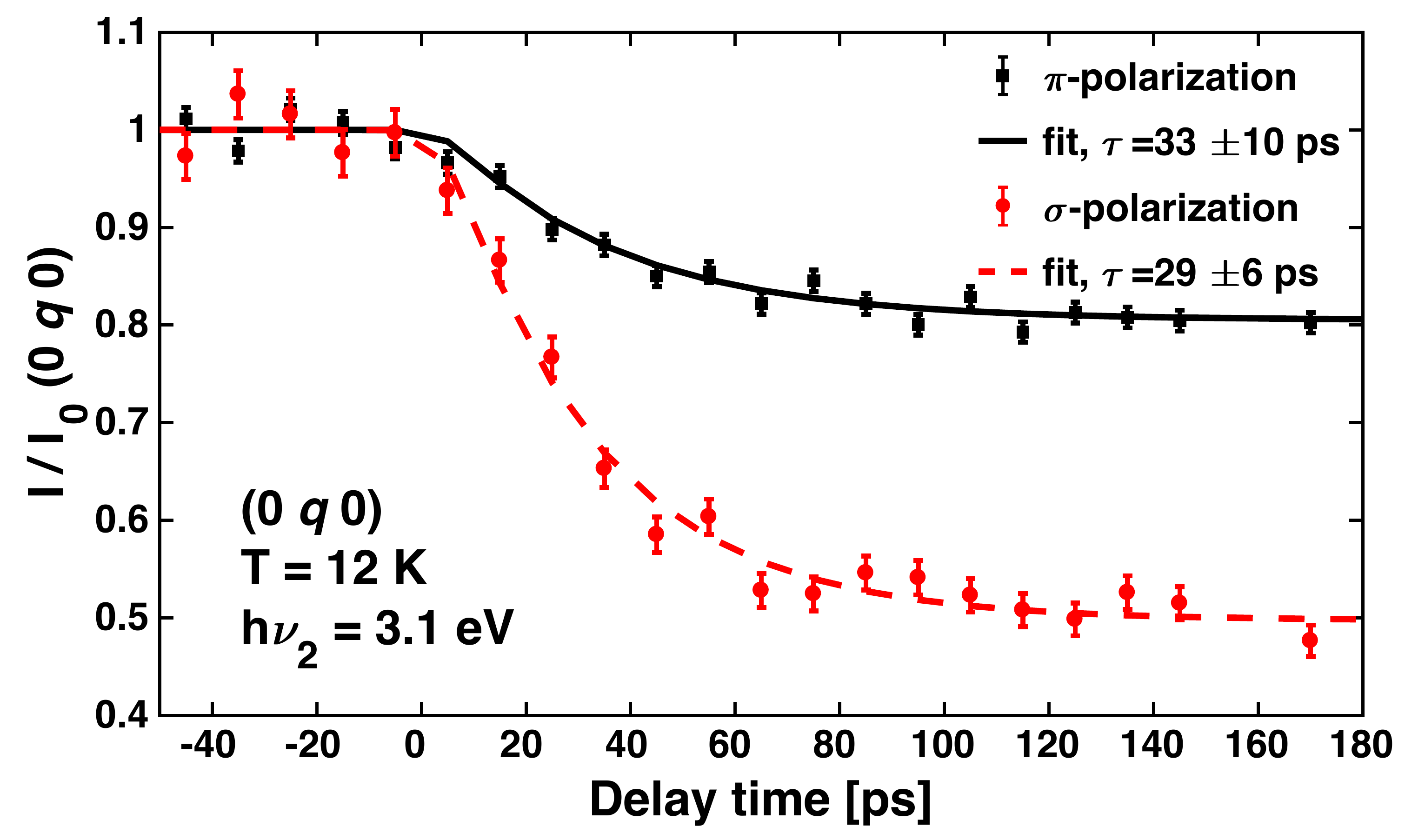}
\caption{
Time-resolved change in diffracted intensity of the (0 $q$ 0) reflection recorded at 12 K at the Mn $L_2$ edge for different x-ray probe polarizations, following photoexcitation at 3.1 eV with 0.8 mJ/cm$^2$ absorbed fluence.
$\pi$-polarization data and fit are shown as black squares and a full black line.
$\sigma$-polarization data and fit are shown as red circles and a dashed red line.
}
\label{q_diffPhEn}
\end{center}
\end{figure}

Fig. \ref{q_diffPhEn} shows the intensity of the (0 $q$ 0) reflection as a function of pump-probe delay time following photoexcitation by 3.1 eV femtosecond pulses.
The (0 $q$ 0)$_\sigma$ intensity drops by 50\% whereas the (0 $q$ 0)$_\pi$ intensity drops by only 20\%, i.e. 2.5 times less.
Decay time constants of 33$\pm$10 ps and 29$\pm$6 ps were extracted for $\pi$- and $\sigma$-polarization, respectively.

Both the timescale and magnitude of the total decrease in diffraction signal for $\pi$- and $\sigma$-polarization following photoexcitation at 3.1 eV exhibit a very similar behavior and are quantitatively quite similar to the results obtained at 1.55 eV, albeit with larger uncertainties because of lower total counts.
Differences in the magnitude of the transient demagnetization signals between the two pump photon energies could be due to slight differences in the spot sizes, which would affect the absorbed energy density, or in the pump induced average heating, which would modify the effective temperature of the sample before photoexcitation.

\subsection{Melting of the orbital order}

\begin{figure} [htb!]
\begin{center}
\includegraphics[width=0.4\textwidth,keepaspectratio=true]{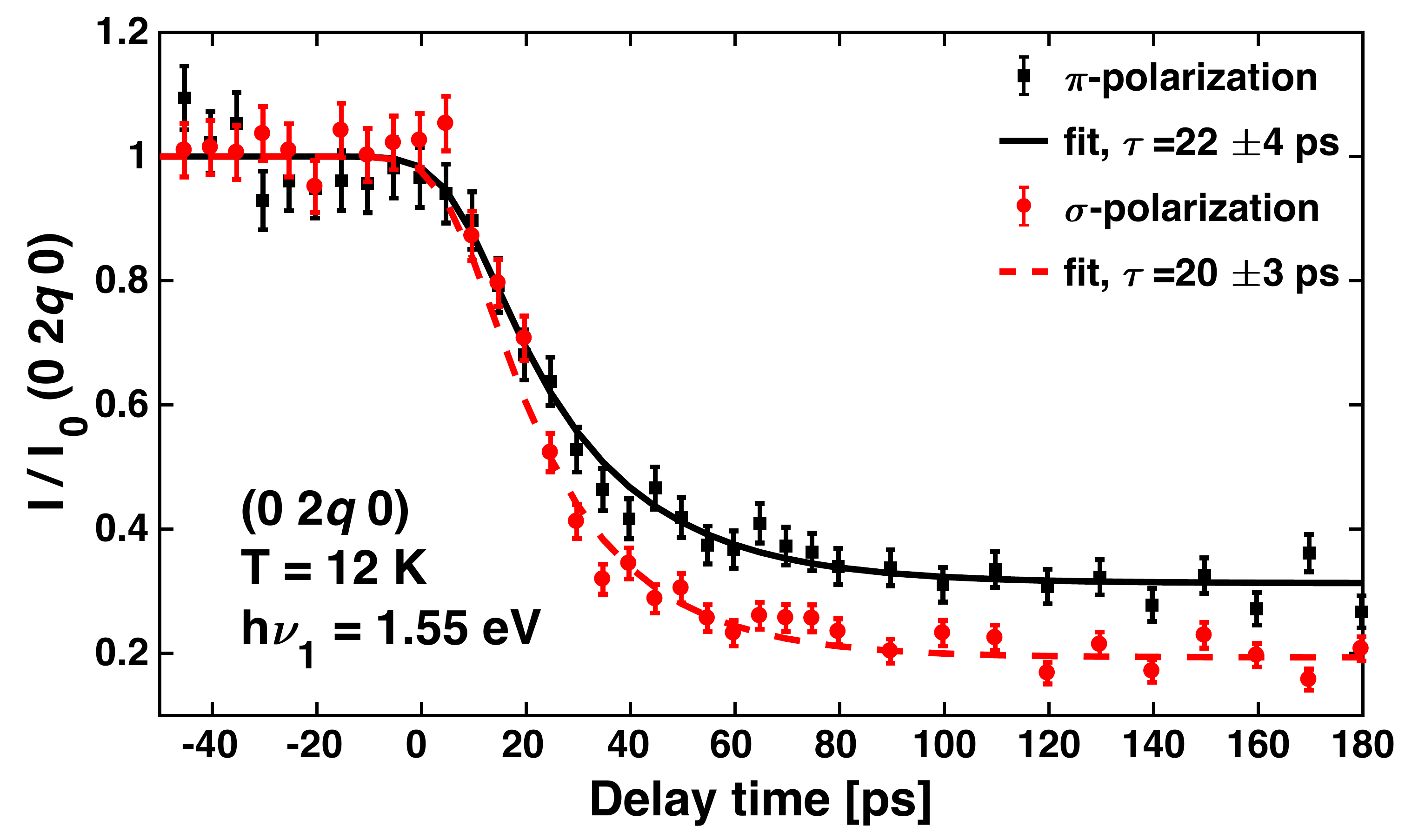}
\caption{
Time-resolved change in diffracted intensity of the (0 2$q$ 0) reflection recorded at 12 K at the Mn $L_2$ edge for different x-ray probe polarizations, following photoexcitation at 1.55 eV with 1.3 mJ/cm$^2$ absorbed fluence.
$\pi$-polarization data and fit are shown as black squares and a full black line.
$\sigma$-polarization data and fit are shown as red circles and a dashed red line.
}
\label{2q}
\end{center}
\end{figure}

In addition to the (0 $q$ 0) response we also analyzed the dynamics of the (0 2$q$ 0) reflection which, as mentioned above, reflects the displacement of atoms due to orbital reconstruction.
The photoinduced change in scattered intensity of (0 2$q$ 0) following excitation by a 1.55 eV femtosecond laser pulse with a fluence of 1.3 mJ/cm$^2$ (absorbed) is shown in Fig. \ref{2q}, measured with a 15.5 ps x-ray pulse duration (FWHM).
Using a single exponential fit, decay time constants of 22$\pm$4 ps and 20$\pm$3 ps can be determined in the (0 2$q$ 0)$_\pi$ and (0 2$q$ 0)$_\sigma$ channels, respectively.

\section{Discussion}

The aim of the present study is to identify the transient state obtained in TbMnO$_3$ following femtosecond laser excitation, as well as to determine the impact of the excitation pathway and of the orbital reconstruction on the demagnetization dynamics.
This knowledge will contribute to determining the origin of the slow magnetic order dynamics in TbMnO$_3$, compared to other antiferromagnetically ordered systems.
As described above, resonant magnetic diffraction at the Mn $L_2$-edge with both $\pi$- and $\sigma$-polarization provides signals specific to the sinusoidal phase, with collinear magnetic order, and to the multiferroic phase, characterized by cycloidal spin order.

Table \ref{table} summarizes the main time constants for the different reflections, polarizations and photon energies, reported in Figs. \ref{q_pisig}-\ref{2q}.

\begin{table}
\centering
\begin{tabular}{ccc}
\textbf{Diffraction condition} & \textbf{$\pi$} & \textbf{$\sigma$} \\
\hline
\hline
(0 $q$ 0), 12 K & \multirow{2}{*}{28$\pm$4 ps} & \multirow{2}{*}{13$\pm$4 ps} \\
h$\nu_1$ = 1.55 eV & & \\ \hline
(0 $q$ 0), 30 K & \multirow{2}{*}{34$\pm$4 ps} & \multirow{2}{*}{---} \\
h$\nu_1$ = 1.55 eV & & \\ \hline
(0 $q$ 0), 12 K & \multirow{2}{*}{33$\pm$10 ps} & \multirow{2}{*}{29$\pm$6 ps} \\
h$\nu_2$ = 3.1 eV & & \\ \hline
(0 2$q$ 0), 12 K & \multirow{2}{*}{22$\pm$4 ps} & \multirow{2}{*}{20$\pm$3 ps} \\
h$\nu_1$ = 1.55 eV & & \\
\end{tabular}
\caption{
Fitted decay times for different diffraction conditions in $\pi$- and $\sigma$-polarization, corresponding to the data in Figs. \ref{q_pisig}-\ref{2q}.
Error bars indicate 95\% confidence intervals.}
\label{table}
\end{table}

First, we consider the magnetic diffraction at the (0 $q$ 0) wave vector after excitation with a photon energy of 1.55 eV.
The 1.8 times stronger reduction in the $\sigma$-channel compared to the $\pi$-channel indicates that the non-collinear magnetic long-range correlations are much more sensitive to ultrafast photoexcitation than the collinear magnetic order.
This is consistent with the scenario in which the laser pulse heats the system along the cycloidal $\rightarrow$ sinusoidal $\rightarrow$ non-magnetic transition path, similar to the adiabatic temperature dependent transition.
As seen in Fig. \ref{q_pisig} and Table \ref{table}, the $\sigma$-polarized (0 $q$ 0) signal decays slightly faster than the $\pi$-polarized signal following 1.55 eV pumping.
Given the very similar slope of the static magnetic order signal vs. temperature curve for $\pi$- and $\sigma$-polarized channels (Fig. \ref{static_q}c), one would expect the dynamics to be the same for both.
The faster timescale exhibited by (0 $q$ 0)$_\sigma$ at 1.55 eV pump is likely caused by a saturation effect due to parts of the excited region transitioning completely into the sinusoidal phase.
In this sense the shorter decay time is an artifact, which originates from strong excitation of the system such that the cycloidal phase is essentially extinguished before the nonequilibrium final state is reached.
This apparent acceleration of the decay time due to saturation is consistent with previously reported measurements of the fluence dependent $\pi$-polarized signal dynamics \cite{Johnson2015}.
Note that the fluence value for which saturation of the $\sigma$-polarized signal is observed in our study cannot be directly compared with those used in Johnson \emph{et al.} \cite{Johnson2015}.
The latter study focused on the $\pi$-polarized signal, which disappears only at T$_{N_1} >$ T$_{N_2}$ (Fig. \ref{static_q}), and had a lower initial temperature, due to lower average heating as discussed above, so that higher fluence values are needed to reach saturation of the dynamics.
Complete transition of parts of the pumped volume is not incompatible with a (0 $q$ 0)$_\sigma$ signal drop of only 70\% (Fig. \ref{q_pisig}) given the above mentioned inhomogeneity of the photoexcited region.

When the base temperature of the sample is increased above T$_{N_2}$, i.e. in the sinusoidal phase where no $\sigma$-polarized signal remains, the decay time in the $\pi$-channel stays similar to the one at low temperature (Fig. \ref{q_12K30K} and Table \ref{table}), for the same relative intensity drop.
This is consistent with the similar slope of the static magnetic order signal vs. temperature curve in the two magnetic phases (Fig. \ref{static_q}c).
All-optical measurements report a clear increase of the decay time with decreasing temperature, especially below T$_{N_1}$ \cite{Handayani2013}.
The transient reflectivity experiment by Handayani \emph{et al.} \cite{Handayani2013} is, however, a measure of the optical properties of the magnetic phase rather than of the magnetic order directly.
Better time resolution would be necessary in order to verify this apparent discrepancy between the observations of optical reflectivity and resonant soft x-ray diffraction.

From the discussion of the (0 $q$ 0) reflection following 1.55 eV excitation and taking into account the fact that the $\pi$-channel ($\sigma$-channel) response is predominantly sensitive to the long-range collinear (non-collinear) magnetic order, we conclude that the magnetic interaction leading to the sinusoidal correlations are stronger than the ones responsible for the cycloidal order so that they are less affected by photoexcitation.
Consequently, the transient magnetic phase that arises after photoexcitation of the multiferroic phase of TbMnO$_3$ exhibits mainly sinusoidal magnetic order, as in the adiabatic case. 

The transient magnetic state we observe therefore resembles the sinusoidal state from the thermal adiabatic transition in terms of its polarization-dependent intensity.
However, Johnson \emph{et al.} \cite{Johnson2015} have observed that the diffraction wave vector of this transient state does not change along with the picosecond scale decrease in intensity following photoexcitation as it does for a slow and adiabatic increase in temperature.
The authors suggest that the change in the diffraction wave vector might be delayed due to a low magnon group velocity, drawing from equivalent arguments based on phonon group velocity which are used to describe the delayed lattice expansion response to ultrashort electronic excitations \cite{Senff2007}.
Our observations confirm that the cycloidal magnetic order melts first and that the transient state is a sinusoidal magnetic phase which, however, retains the wave vector of the low-temperature multiferroic phase.

Alternatively, our observations could be accounted for by a photoinduced flop of the magnetic cycloid from the bc- to the ab-plane.
An ultrafast electronic excitation will likely affect the equilibrium between different magnetic contributions, namely between superexchange and the Dzyaloshinski-Moriya interaction.
Modifying the relative contribution of these interactions could favor an ab-plane cycloid over the usual bc-plane cycloid of TbMnO$_3$. \cite{Mochizuki2009}
This would cause the $\sigma$-polarized response to disappear and the $\pi$-polarized signal to decrease, while preserving the wave vector of the bc-cycloid reflection.
While our data do not invalidate this alternative scenario they also provide no indication that it is more likely than our first, simpler interpretation.
We therefore do not take it into account in the rest of the discussion. 

Next, we compare the results obtained for the two different excitation photon energies, 1.55 eV (exciting mainly Mn $3d$ - Mn $3d$ excitations) and 3.1 eV (exciting mainly O $2p$ - Mn $3d$ excitations), at approximately equivalent values of the absorbed excitation density.
Similar behavior is observed for the two photon energies: the $\sigma$-polarized signal decreases more than the $\pi$-polarized signal, and the decay timescales are both on the order of 30 ps (Fig. \ref{q_diffPhEn} and Table \ref{table}).
When comparing the 3.1 eV excitation (Fig. \ref{q_diffPhEn}) with the excitation at 1.55 eV (Fig. \ref{q_pisig}), the overall drop in intensity is slightly larger for the 1.55 eV excitation.
We attribute this to systematic errors in estimating the fluence of the 3.1 eV excitation needed to match the absorbed energy density for the 1.55 eV excitation, which is based on literature values for the absorption coefficients \cite{Bastjan2008}.
The fact that the $\sigma$-polarized signal with 3.1 eV excitation decays at the same rate as the $\pi$-polarized one, and not faster as with 1.55 eV excitation, is consistent with the lack of saturation effect expected for a smaller transient decrease of (0 $q$ 0)$_\sigma$.
Some of the small quantitative differences between the dynamics at the two excitation photon energies could also arise from differences in the average heating caused by the two pump pulses.

The comparison between 1.55 eV and 3.1 eV photoexcitation leads us to the conclusion that the excitation photon energy and therefore the nature of the electronic excitation from either Mn $3d$ or O $2p$ states does not play a significant role in determining the timescale of long range magnetic order melting following photoexcitation.
That is, the slow demagnetization behavior in TbMnO$_3$ cannot be explained by some particularity of the excitation path following a 1.55 eV pump.
The energy density absorbed by the system, which eventually leads to heating of the lattice, seems to be the most relevant parameter.

We now discuss the dynamics of the orbital reconstruction via the (0 2$q$ 0) reflection response to excitation at 1.55 eV.
As seen in Fig. \ref{2q}, both polarization channels show a decrease in diffracted intensity by roughly the same relative amount, which is fully compatible with the picture of a quasi-adiabatic heating of the system from the cycloidal to the sinusoidal phase (Fig. \ref{static_2q}c).
The slightly larger transient intensity drop in (0 2$q$ 0)$_\sigma$ (~80\%) compared to (0 $q$ 0)$_\pi$ (~70\%) can probably be explained by a slight variation in pump fluence and average heating, due to the difference in incidence angle for the two reflections, as well as by a difference in the resonant absorption for the two reflections, leading to different penetration depths of the soft x-ray probe pulses.
The slightly larger transient intensity drop in (0 2$q$ 0)$_\sigma$ (~80\%) compared to (0 $q$ 0)$_\sigma$ (~70\%) is consistent with the different temperature dependence exhibited by (0 $q$ 0) (Fig. \ref{static_q}c) and (0 2$q$ 0) (Fig. \ref{static_2q}d).
This different intensity drop could also be related to a slight variation in pump fluence and average heating, due to the difference in incidence angle for the two reflections, as well as to a difference in the resonant absorption for the two reflections, leading to different penetration depths of the soft x-ray probe pulses.
We observe that the temperature-dependent (0 2$q$ 0) intensity decreases significantly but does not completely disappear when going from the cycloidal to the sinusoidal phase (Fig. \ref{static_2q}d), a behavior in between that of the $\pi$- and $\sigma$-polarized channels for (0 $q$ 0) (Fig. \ref{static_q}c).
Assuming coupling between magnetic and orbital ordering, we would therefore expect the decay timescale associated with the photoexcited (0 2$q$ 0) reflection to accompany that of (0 $q$ 0).
This is exactly what is seen in our data (Fig. \ref{2q} and Table \ref{table}).

Similar timescales between (0 $q$ 0) and (0 2$q$ 0) indicate that magnetic and orbital order are coupled, meaning that there is no evidence that photoexcitation affects the electronic orbital reconstruction directly even though the pump pulse directly excites the electronic ground state of TbMnO$_3$.
This is contrary to what happens in other systems, such as in half doped manganites where a 1.55 eV excitation promptly melts the orbital order \cite{Beaud2014}.
The difference might be due to the different origins of the two orbital reconstructions: in TbMnO$_3$ it arises from the cycloidal magnetic order, while in half doped manganites the orbital order arises directly from electronic correlations \cite{Dagotto2001}.

Johnson \emph{et al.} \cite{Johnson2015} interpreted the dynamics of the magnetic diffraction from TbMnO$_3$ in terms of an effective time-dependent spin temperature.
The stronger influence of average heating in our experiment prevents an accurate determination of the initial temperature, which in turn hinders a precise estimate of the transient spin temperature following photoexcitation.
We note, however, that given the similarity of the observations between our experiments and that of Johnson \emph{et al.} \cite{Johnson2015} we would expect comparable results from applying the same time-dependent spin temperature analysis to our data, both for the (0 $q$ 0) and for the (0 2$q$ 0) reflections.

Finally, we discuss why the demagnetization dynamics are slow in the multiferroic phase of TbMnO$_3$ relative to other transition metal oxide antiferromagnets.
All our present observations seem to indicate that ultrafast photoinduced demagnetization follows the adiabatic thermal pathway measured statically.
Heating of the electronic subsystem occurs, however, in a few hundred femtoseconds, so that the dynamic bottleneck for the demagnetization process must be due to the lattice or to the spin system itself, or to a combination of both.
A complete microscopic description of the mechanism remains nevertheless lacking. 

We can gain some insights into the possible microscopic mechanism by considering recent model calculations performed in the framework of nonequilibrium DMFT \cite{Aoki2014}, which enables the simulation of ultrafast electronic relaxation and demagnetization dynamics in highly excited correlated electron systems.
We consider, for simplicity, a single-band Hubbard model in the Mott insulating regime.
For this system, it has been shown that the lifetime of photoexcited doublons (doubly occupied sites) and holes depends exponentially on the interaction strength or gap size, and can thus easily reach picosecond timescales for realistic parameters \cite{Eckstein2011}.
If the Mott insulator is antiferromagnetically ordered, and the inital kinetic energy of the photoexcited carriers is too low to disorder the spin background, the system is trapped in a long lived magnetized state whose lifetime is controlled by the doublon-hole recombination time. \cite{Werner2012}
Indeed, after the initial relaxation within the Hubbard bands, the doublons and holes have very low kinetic energy, and most of the energy injected by the pulse is stored as potential energy.
The latter is slowly released through doublon-hole recombination processes, and is transfered via an increased kinetic energy of the remaining carriers to the spin background.
Considering the analogy between the Hubbard band structure of the model and the Mott-induced $d$-$d$ splitting in TbMnO$_3$ \cite{Moskvin2010}, the observed slow demagnetization dynamics could thus be understood as a consequence of an electronically trapped magnetic order resulting from limited kinetic energy of the photodoped carriers and slow doublon-hole recombination.

In real materials, however, lattice heating and spin-phonon scattering will become relevant on picosecond timescales. 
For a quantitative comparison, extensions of the theory to multi-orbital systems \cite{Strand2017} and electron-phonon coupling (Holstein-Hubbard Model) \cite{Werner2013} are required.
Our experimental observations can serve as a test for such more realistic calculations.

\section{Summary}

We have studied the demagnetization dynamics of the cycloidal and sinusoidal magnetic order in multiferroic TbMnO$_3$ by means of time-resolved optical pump – resonant soft x-ray diffraction probe.
Using orthogonal linear x-ray polarizations, we are able to distinguish the responses of the multiferroic cycloidal spin order and the sinusoidal antiferromagnetic order in the time domain.
We show that magnetic order suppression following intense photoexcitation of the electronic system occurs via a sinusoidal transient phase that arises on a timescale of tens of picoseconds.
This intermediate phase is similar to the one observed during the adiabatic temperature-dependent transition but retains the low temperature wave vector.
Using two different pump photon energies, 1.55 eV and 3.1 eV, the conduction band is predominantly populated by intersite Mn $3d$ - Mn $3d$ transitions or intrasite O $2p$ - Mn $3d$  transitions, respectively.
We find that the nature of the optical excitation does not play an important role in the melting of magnetic order as far as the timescale of the energy transfer between the electronic and the spin system is concerned.
We further observe the reduction of orbital reconstruction on a timescale comparable to that of cycloidal order melting, which suggests that the orbital reconstruction is not directly affected by the photoexcitation.

Our work contributes to a better understanding of the ultrafast demagnetization pathways in photoexcited TbMnO$_3$.
Further investigation is needed in order to determine the microscopic mechanism that drives this magnetic order melting, both experimentally, for instance by measuring the lattice dynamics following photoexcitation, and theoretically, by extending existing models to multi-orbital systems and including electron-phonon coupling.

\section{ACKNOWLEDGEMENTS}

The research leading to these results has received funding from the Swiss National Science Foundation and its National Center of Competence in Research, Molecular Ultrafast Science and Technology (NCCR MUST) and Materials' Revolution: Computational Design and Discovery of Novel Materials (NCCR MARVEL).
E. M. B. acknowledges funding from the European Community's Seventh Framework Programme (FP7/2007-2013) under grant agreement n.°290605 (PSI-FELLOW/COFUND).
E. A. acknowledges support from the ETH Zurich Postdoctoral Fellowship Program and from the Marie Curie Actions for People COFUND Program.
Crystal growth work at IQM was supported by DOE, Office of Basic Energy Sciences, Division of Materials Sciences and Engineering under Award DE-FG02-08ER46544.​
M. F., Y. W. W., and M. R. acknowledges support by SNSF project 200021\_147080, 137657, and CRSII2\_147606, respectively. 

This work was carried out with the support of the Diamond Light Source.
We thank Diamond Light Source for access to beamline I06 (proposal numbers SI 13355 and SI 13926) that contributed to the results presented here and thankfully acknowledge support by the beamline scientists Y. Liu and F. Maccherozzi.
We thank Ekaterina Pomjakushina of the Laboratory for Scientific Developments and Novel Materials and Stefan Stutz of the Laboratory for Micro- and Nanotechnology at the Paul Scherrer Institute for help with the preparation of the sample.
We acknowledge the Paul Scherrer Institute, Villigen, Switzerland for provision of synchrotron radiation beamtime at beamline X11MA (SIM) of the SLS and would like to thank the X11MA beamline staff for assistance.


\bibliographystyle{apsrev4-1}
\bibliography{TbMnO3Diamond}

\end{document}